\documentclass[aps,prb,preprint,noshowpacs,showkeys,amsmath]{revtex4-1}

\usepackage{amsmath}
\usepackage{amssymb}
\usepackage{amsfonts}

\begin{document}

\title{How to introduce physical quantities physically}

\author{Serge A. Wagner}
\email{s\_wagner@mail.ru}
\affiliation{Theoretical Physics Department, Moscow Institute of Physics and Technology, Dolgoprudny, Moscow region 141700}

\date{\today}

\begin{abstract}
The centuries-long practice of the teaching turned mechanics into an academic construct detached from its underlying science, the physics of macroscopic bodies. In particular, the regularities that delineate the scope of validity of Newtonian mechanics was never used as premises in that construct. Instead, its logical structure has been built with the only purpose to ease the presentation of mechanics as an application of calculus and algebra. This leaves no room for the explicit physical description of the fundamental notions of the construct, such as the ability of (a system of) physical bodies to keep a spatial form, symbolized by a rigid measuring rod, and the possibility to count out even temporal intervals, symbolized by a standard clock, let alone the origin of the basic mechanical quantities.

The comparison between (states of) physical objects is possible as far as the natural regularities enable a researcher to define an equivalence relation in a set of these objects. The class of macroscopic (or macroscopically perceived) objects is that one where certain composition rules hold so that constituents of any object are macroscopic, too. Then the properties of compositions of macroscopic objects, being in conjunction with the equivalence relations between them, yield a numerical characterization of each macroscopic object, i.e. a basic physical quantity. The technique of logically arranged feasible/thought experiments is used to formulate the logically impeccable, induction-based definitions for two quantities: the length of a line segment as an instructive toy example and the mass of a macroscopic body as an example applicable to the real world.
\end{abstract}

\maketitle

\section{Is physics as foundationless as it is presented?}\label{foundationless}
 
It is commonplace for standard physics courses to introduce the basic physical quantities, such as length, time, mass, charge, force, temperature, in a fairly artless, not to say misleading, manner. Effectively, they provide students with nothing but a bare reference to the symbols of the mathematical variables that can derive their ultimate theoretical and experimental meaning from the topics covered later on in the curriculum. Obviously, this instigates a student to begin learning physics with rote learning since ordinarily we expect symbols to refer to things or events or ideas, but not the reverse. An educated person, such as a teacher, perceives this organization of the learning material as logically circular; it completely neglects those subtle and careful considerations that the scientists of the past had to address when developing the basic concepts gradually.\cite{original presentations} This can easily make inexperienced teachers feel involved in expounding a foundationless and thereby questionable science.

The practice of teaching impels authors of textbooks, especially those of the high school level, to moderate and to disguise the apparently awkward, upside-down logic of the introductory chapters. Usually they try to appeal to students' everyday experience, such as perception of muscular effort in exemplifying the concept of the force as a push or a pull, sense of hot and cold in presenting the idea of the temperature as a degree of warmth, vague reminiscences of the obsolete meanings of the mass as a quantity of matter and the charge as a quantity of electricity. But eventually, remarks of this kind either give way to common misconceptions and make the long-term distraction from the precise meaning of the physics terms or, at best, result in a waste of time because no correct logical inference or calculation uses them actually. Quite aware of this, most of the authors of university level texts have opted to restrict each such initial remark to a one-sentence opening statement and to start the algebra and calculus based presentation of physical laws as early as possible. However, without the essential and well-founded definitions of physical quantities,\cite{casual definitions} any formulas remain largely meaningless until the students set about working through examples and thereby get access to the experimentally verifiable information such as the \textit{non-numerical} but unambiguous description of situations where a given formulation of a given law is applicable. 

In the next section, the reader will have a chance to appreciate anew how incredibly long, inevitably vague, and eventually useless the chain of those "introductory" statements can be, with which the authors of modern textbooks feel obliged to precede the analysis of the more or less true physical examples.

\section{Is science perplexingly simple or simply perplexing?}\label{perplexing science} 

Although the perfunctory presentation of the primary theoretical concepts in standard textbooks may seem sufficient to start making students acquainted with the collection of formulas prerequisite to training in engineering disciplines, it is obviously far from expounding physics as a science. Moreover, attempts to adhere to the logical sequence within the approach sketched in the previous section could even discredit the scientific method because they would formally entail the ludicrous necessity to involve complicated and hazy notions in order to interpret simple and clear ones. The following is an example illustrating this point.

Evidently, the regular use of kit\-chen/bath\-room/lug\-gage scales allows observant people to get the idea how some of their readings are related, without any theoretical reasoning. In fact, it is elementary school science lessons that are supposed to help virtually all students gain this kind of experience. One class of such relations expresses the result
\begin{equation}\label{additivity of the weights}
  W[a\cup b]=W[a]+W[b]
\end{equation}
of weighing two body \(a\) and \(b\) simultaneously in one scalepan via those of weighing them separately. If, additionally, one takes into account the measurement procedures, one can further make a distinction between Eq. \eqref{additivity of the weights} and a similar property of bodies themselves,
\begin{equation}\label{additivity of the mass}
  m[a\cup b]=m[a]+m[b],
\end{equation}
produced by weighing with the aid of a beam balance and commonly referred to as the additivity of \textit{mass}. Note that the \textit{form} of Eq. \eqref{additivity of the weights} or Eq. \eqref{additivity of the mass} reveals the exact meaning of the word "quantity" while the actual \textit{physical operations} denoted by the symbols "\(\cup\)" and "\(+\)" specify what sort of quantity we are dealing with. (Those who still can't tell much difference between a mathematical number and a quantity of something might find it enlightening to look at the appendix \ref{subtraction examples}.)

Over centuries, the facts expressed by Eq. \eqref{additivity of the mass} motivated students to adopt "a quantity of matter" as a definition of mass willingly, assuming that "matter" is an indestructible \textit{ponderable} entity able to occupy space as well as actually or potentially divisible. Aside from this extraneously speculative and unnecessarily generalizing conception, the physical regularities backing up Eq. \eqref{additivity of the mass} enable the practical standards of mass known as prototypes and still being in effect.

Similarly, Eq. \eqref{additivity of the weights} exposes the weight \(W\) as another quantity. It allowed teachers of the past to use the weight as an easily produced and readily available exemplar of the force and then to unfold statics consistently, without invoking Newton's second law of motion prematurely.\cite{old presentations}

By contrast, the modern physics education does not provide students with an orderly scientific description of what they can gradually learn about nature. Had today's learner been resolute enough to extract it as completely as possible from the common modern textbooks how physics treats his/her plain experience of weighing, he/she would have to work hard in order to finally build something like the following long sequence of statements:
\begin{enumerate}
   \item\label{definition of position} The position of a body assumed as a material point is (represented by) a vector whose components in a reference frame aka a coordinate system are the Cartesian coordinates of the body.\cite{presentation of Cartesian coordinates} Coordinates are assumed to be numbers resulting from the measurement with the aid of a meterstick (a measuring rod) or a measuring tape.\cite{existence of coordinates}
   \item\label{definition of time} The positions of a bodies may change in the course of time. The time interval is measured by properly synchronized clocks in an inertial reference frame.\cite{existence of time}
   \item\label{definition of velocity} The velocity of a body is the first derivative of its position vector with respect to time; the acceleration \(\mathbf{a}\) of a body is the first derivative of its velocity with respect to time.\cite{time derivatives}
   \item\label{definition of force} The cause of the acceleration of one body is an action of other bodies. The standard action/force aka the unit of force is an effort (a push or a pull) producing a standard acceleration of a standard body.\cite{standard force} Then the force of a given action is supposed to be a number of the standard ones in the same direction and together making the given action.
   \item\label{alternative definition of force} Any push or pull action has its own characteristics: The masses of two point bodies and the relative position vector of one body with respect to another determine the universal gravitation between them uniquely; the similar statement holds for the electrical interaction of two point electric charges; a specific compression of a spring dictates its stress etc. So, in principle, we can make an alternative choice of a standard force by singling out one with some specified characteristics, for example, the tension in a certain piece of rubber stretched to a certain extent.\cite{alternative standard force} The spring scale calibrated against the weight of a standard body can be used to measure other kinds of forces.\cite{measurement of force}
   \item\label{superposition of forces} Experiments show that several forces simultaneously acting on a body combine to make a net force in accordance with the vector addition rule.\cite{force combination}
   \item\label{definition of an inertial frame} An inertial reference frame is one where the inertia law aka Newton's first law holds.\cite{criterion for inertial frame} In another formulation, this is a frame where Newton's laws together stand.\cite{another criterion for inertial frame}
   \item\label{definition of the mass} In an inertial reference frame, a net force \(\mathbf{F}\) acting on a given body effects the body's acceleration \(\mathbf{a}\propto\mathbf{F}\) so that the relation \(F/a\) depends on the body only and, for this reason, is called the mass of the body.\cite{definition of mass} Then Eq. \eqref{additivity of the mass} appears consistent with the application of Newton's second law to a compound of two rigidly connected bodies.\cite{additivity of mass}
   \item Alternatively, the relation of accelerations of interacting bodies can yield the relation of their masses. Thus the adoption of a certain number as the mass of a standard body results in assigning definite numbers to the masses of all bodies actually or potentially.\cite{alternative definition of mass} Employing this definition, one can experimentally verify that "mass is a scalar quantity and thus obeys the rules of ordinary arithmetic." \cite{scalar quantity}
   \item\label{definition of the weight} The weight \(W\) of a body of mass \(m\) equals the magnitude \(mg\) of the gravity force where \(g\) is the magnitude of the acceleration of a body freely falling near the Earth's surface.\cite{definition of weight}
\end{enumerate}

Consider a university student versed enough to understand at least the formal relations among the quantities mentioned in the above items. When asked to explain/interpret Eq. \eqref{additivity of the mass} on the basis of physical laws, such a student would not find it difficult to see that, in view of the simple chain of equalities,
\[m[a]\cdot g+m[b]\cdot g=W[a]+W[b]=W[a\cup b]=m[a\cup b]\cdot g,\]
the items \ref{definition of the weight}, \ref{superposition of forces} and \ref{definition of the mass} entail Eq. \eqref{additivity of the mass}. Meanwhile, an experienced physicist would find this reasoning at least insufficient and perhaps as physically pointless as, for instance, the naive formal derivation of the inertia law from Newton's second law applied to a free body. (For another instance of the obvious but naive "inference", see the discussion after Eq. \eqref{segment commutation} in Section \ref{addition}.)
This is because drawing conclusions about such basic quantities from the general physical regularities requires a profound knowledge of their true meaning. But, keeping in mind a supposedly elaborated and authoritative sequence like the items \ref{definition of position} - \ref{definition of the weight}, the beginner would hardly doubt his/her "theory" about weighing even when he/she came across isolated warnings such as "this additive property of mass seems obvious but it has to be verified experimentally." \cite{verification} The student would simply not make out such comments since they dismiss his/her attempt to rely upon deduction,  i.e. the idea that a conclusion is as true as the premises and needs no separate testing.

In the next section, the reader will learn why the presentation of mechanics in today's textbooks has no choice but to tolerate the false introduction to the basics.

\section{Can science be logical?}\label{logic of science}

The apparent disregard for the logic -- the main tool of theoretical analysis -- in the warning quoted in the end of the previous section is only a tip of the iceberg: The sequence \ref{definition of position} - \ref{definition of the weight} makes no scientifically reasonable and logically ordered (basis for a) body of knowledge about a particular field of physics but is the historically established way of introduction to the set of the regularities known as Newtonian mechanics, by means of algebra and calculus.\cite{culture} Further on in the essay this sequence is referred to as the academic construct of mechanics, or AM in the abbreviated form.

If you acquaint students with the formalism of mechanics in order to start applying algebra/calculus to a host of hopefully instructive examples/models as early as possible, then it does not matter whether you stick to an overall consistent view or not. Some minor but obvious departures from the logical order in the academic construct of mechanics are just slightly annoying. For example, its items \ref{definition of force} and \ref{alternative definition of force} imply that the addition operation for forces has already been defined while in fact the item \ref{superposition of forces} of AM describes it later. This is tolerable in a general/overview course prerequisite to major engineering subjects, when a teacher focuses on solving idealized problems that symbolize feasible applications rather than represent fundamentals of nature.

Meanwhile, students studying physics as a science, especially as a major subject, are usually expected to be good at critical thinking and rational reasoning. So the subject they take ought to be logically consistent at least. In this relation, the drawbacks of the academic construct of mechanics are too awkward to fix without restructuring the entire presentation. If learners started questioning, for example, the traditional wording of the inertia law, a teacher could not give them a logically correct explanation within the frame of that construct.\cite{Rigden1987} The following are the remarks about the most essential flaws in the academic construct of mechanics.

One challenge for the presentation of physics theory comes from the continuous refinement of mathematics education: It is not uncommon today for a university physics teacher to deal with physics/mathematics majors with a strong mathematical background, who can be interested in why physical quantities are real numbers, not rational or hyperreal ones, or even elements of a non-commutative algebra. At first glance, it is only an introduction to an operator formalism in a graduate/postgraduate course of quantum mechanics that can provoke such questions. However, as soon as the learner understands that the concept of numbers of a given sort is primarily defined by specifying the addition and multiplication operations with them, he/she can clearly comprehend what the additivity of mass or the superposition of forces means and may guess that some physics should underlie the addition/subtraction operation for any quantity. Then nothing can prevent such students from questioning, for instance, even the mundane and seemingly unassailable item \ref{definition of velocity} of the academic construct of mechanics, since the derivative involves subtraction and, therefore, should imply some way to make it physically. (Readers with little experience in relation of mathematical constructs such as addition/subtraction to their applications may find the appendix \ref{subtraction examples} elucidative.)

One of the more subtle but apparent formal faults in the academic construct of mechanics is that the definition of a time \textit{coordinate} in the item \ref{definition of time} of AM employs the concept of an inertial reference frame, which, in turn, refer to Newton's laws in the item \ref{definition of an inertial frame} of AM. It reveals that in fact dynamics precedes kinematics logically, which is contrary to the usual presentation order of the AM items \ref{definition of velocity} and \ref{definition of force}. 

The fundamental flaw that undermines the logic of the academic construct of mechanics from the beginning is as follows: The item \ref{definition of position} of AM implies that coordinates and, therefore, a reference frame itself \textit{exist} physically (i.e. can be characterized as measur\-able/ob\-servable/realizable/reproducible), only if it is possible to independently verify whether the measuring rod is rigid and applied consistently with any other rigid objects.
Similarly, the measurement of the time intervals in the item \ref{definition of time} of AM presupposes that the clock ticks evenly or, equivalently, there is no slow uncontrolled variation in a standard frequency. In principle, the conditions for existence of the spatial and temporal coordinates can be coupled when one brings into play electromagnetic radiation, (recognized as) propagating at a constant speed through the vacuum. But if one is wise enough to avoid muddling essentially different phenomena in one's theoretical analysis without a real need and to stay within the mechanics of macroscopic bodies only, one will admit that "rigid" should mean something like "stiff enough under typically applied loads." This suggests that notions from \textit{statics} do take logical precedence over such \textit{geometrical} concepts as Cartesian coordinates.

It is worth emphasizing that for the mechanics of macroscopic bodies and for celestial mechanics the situation described reveals a badly formulated but still working theory. However, any mindless attempt to go further and to apply the same approach to the phenomena at atomic or cosmological scales includes the risk of getting unusable mixture of true, false and meaningless statements because, naturally, we cannot rely on solid rigid rod to interpret the meaning of microscopic or cosmic distances.

In today's education, even a physics major has little chance to get any clarifications on the above and other issues related to the foundations of classical mechanics: Lecturers of undergraduate courses feel it is their duty to consider the fundamentals in more detail\cite{fundamentals for undergraduates} but never touch the origin of the conventional mathematical formulation in their disparate comments. Leaders of graduate courses, especially those interested in foundations of mechanics, may sometimes recognize the logical gaps in its formalism\cite{foundation for graduates} but do not expect them to affect any practical applications. So a student can easily find several mathematical reformulations of classical mechanics such as Lagrangian mechanics, useful for, e.g., computational robotics, but never learns about the logical hierarchy of physical concepts. And for advanced postgraduate courses, typically tailored to meet the immediate needs of research programs, it seems almost impossible to pay attention to such scientifically "old-fashioned" topic as the basic mechanics of macroscopic bodies.

One result of such training is a researcher who has difficulty in formulating the logically complete description of more or less directly perceivable macroscopical physics but is nevertheless supposed to consciously apply more abstruse concepts of quantum mechanics and general relativity to the phenomena on atomic and cosmological scales. It is not uncommon when a young physicist, especially one engaged in teaching, feels as though, in science, there is no firm ground to stand on. And as an instructor he/she usually see no option but to maintain the vicious circle of obscuring the physical origin of the basic quantities for the next generation of students.

In principle, a researcher is assumed capable of self-learning through the study and critical analysis of the works of all previous scholars. But when he/she delves as deep as into the foundations, this kind of his/her individual activity becomes highly time-consuming. So it appears to be nothing but the initial training that generally determines the performance of researchers in their most productive years. Then we have to recognize that, independently of the amount of the government grants for research programs, only the efforts of teachers in developing appropriate curricula maintain the overall quality of research. Although in today's society there seems to be no one interested in supporting such efforts as yet, we should hope that teachers themselves feel obliged to form the consistent and well structured body of knowledge before imparting it.

In the next section the reader will learn what it is in the good old science of the past that can help teachers improve the logic of introductory physics courses.

\section{Are good ideas of old-fashioned science still biding its time?}\label{old science}

In the broad historical sense, the scientific activity is the attempts to alternate the following two processes:
\begin{description}
   \item[linking]\label{science activity} establishing new patterns in available ob\-ser\-va\-tional/experimental/practical data to im\-prove/change/create ideas/hy\-poth\-e\-ses/theo\-ries/technologies;\cite{amalgamation} 
   \item[accumulating] utilizing available tech\-nol\-o\-gies/theo\-ries/hy\-poth\-e\-ses/ideas to obtain new obser\-va\-tion\-al/ex\-per\-i\-men\-tal/practical data.
\end{description}
These processes differ by the type of inference used by researchers: induction is the leading technique in the first process, while deduction is the main tool in the second one.

The considerations in the previous sections suggest that, as far as research activity is concerned, the contemporary physics education does well in preparing physics students for the process of accumulating while its little attention to the process of linking accounts for the existing inconsistencies in the presentations of fundamentals. Meanwhile, many scientists of the past gave us a good deal of reasoning that are well within that latter process. Even Newton, who started his \textit{Principia} with "definitions" and "axioms, or laws of motion" in apparent attempt to follow the deductive pattern of famous Euclid's \textit{Elements}, had recourse to induction and could not but admit it in his Rule IV\cite{Newton Principia: Rules} as a necessary part of reasoning in science. Moreover, only induction can substantiate many of the assumptions on which deductive lines of thought in physics often rely implicitly.

One of the relevant examples in \textit{Principia} is the well known reasoning\cite{Newton Principia: Axioms} by which Newton extends the scope of applicability of his third Law of motion to include long-range attractions such as gravity or magnetism. Thanks to the studies of collisions of macroscopic bodies made by previous authors, there had already been the notion of how action and counteraction of contact repulsion (due to resistance of material) of the bodies relate. Newton took it for granted that there could always be an obstacle that impedes contact between two mutually attracting bodies while maintaining its state with respect to those bodies.\label{III Law} He backed up his idea by reference to an experiment with two magnets and to the apparent equilibrium of the Earth as a self-gravitating body. This suggests that, if stated explicitly, the only reasonable purpose (and the meaning) of Newton's implicit premise would be to delineate the scope of validity of the proposed theory by specifying a class of covered phenomena: the theory is correct as far as there can be an obstacle described by Newton. In particular, it is from the existence of such an auxiliary macroscopic body and from the appropriate empirical data today's physicist can infer, e.g., the lower boundary of the range of the spatial scales within which Newtonian mechanics is applicable (see the appendix \ref{limitation}).

This tacit reference to the experimental/observational data, i.e. use of induction, is typical of such theoretical constructs, later named "thought experiments" (for an overview, see, e.g., \onlinecite{Galili2009} and the references therein.) It is the technique of thought experiments that reseachers/teachers/engineers of the past can invoke to make a non-numeric yet accurate description of a physical situation. Stevin's "principle of solidification"\cite{solidification principle} is a good example of this in mechanics and mechanical engineering.

One of the things that are still important today is that some of such non-numeric statements may remain valid in case the (continuous functions of) Cartesian coordinates of small parts of a body do not exist within a desired accuracy due to, e.g., quantum phenomena at sufficiently small scales. The impossibility of an isolated perpetual motion machine is the most famous principle of this kind: While typically the textbooks tell the learners that this assertion is equivalent to the law of conservation of energy, it in fact is true whether or not there is indeed the energy as a numerical function of the state of motion.

Many works of the past where authors have ingeniously exploited the law of the lever, properties of equilibrium on the inclined plane, the principle of virtual displacements along with the original, synthetic, formulation of Euclidean geometry illustrate that the technique of thought experiments had long been the only tool for the accurate theoretical analysis until the sophisticated algebra and calculus formalism was properly developed. Moreover, since in Euclid's \textit{Elements} there are proofs that make use of such "thought experiments" as rigid motion of geometrical figures, even that work appears to be more related to the theoretical physics (of shapes of solids) rather than to pure mathematics.

The successful use (along with the further development) of calculus in celestial mechanics and continuum mechanics during the 18th and 19th centuries led researchers to its overvaluation and to the perception of reasoning in the form of thought experiments and by means of synthetic geometry as cumbersome. The thought that more complicated argumentation may stem from weaker (i.e. less restrictive) premises was of little value for direct engineering applications. Meanwhile, in search for foundations of geometry, algebra, and calculus, mathematicians of 19th and 20th centuries had developed some toolsets of the low-level analysis such as the set theory and the category theory. However, physics theorists of that time were not prepared to resort to so subtle but abstract tools because they saw no profit in refining the non-numerical description of physical phenomena. Instead, the presentations of Newtonian mechanics as an application of algebra and calculus became dominant in teaching. As a result, invoking thought experiments in the physics literature since the mid-20th century faded into preliminary heuristic reasoning or even mere interpretation of calculus-based calculations. Only a few authoritative physicists have kept the tradition and addressed old known thought experiments as well as invented new ones in their educational activity. The inference of the existence of the gravitational energy of a body (along with the derivation of the well known formula for it) from the impossibility of a perpetual motion machine\cite{gravitational energy} is one of the quite convincing examples of this kind. Still, such cases in point have been too isolated to make a science-based logical structure able to replace the academic construct of mechanics. The apparent cause for this is that non-numerical reasoning in physics is commonly perceived as insufficiently rigorous.

The deductive presentation of Newtonian mechanics established in teaching practice has become a model for the theoretical formulations of all other branches of physics. However, in case of theories of modern physics, such as quantum mechanics and general relativity, the results of this approach appeared shocking for experimenters, since these theories use the quantities that have names familiar from classical physics but not quite the same meaning. While trying to perceive this fact, P. W. Bridgman came up with the idea to define a physical quantity with an aid of operations used in its measurement (see p.~5 and p.~12 in Ref.~\onlinecite{Bridgman1958}). Though provisional and theoretically naive, this idea can also be regarded as an attempt to fill the logical gaps arisen at the foundation of the theory due to downgrading thought experiments.

In the next section the reader will learn of those pre-numeric concepts which cannot be avoided in the rigorous introduction of a basic physical quantity within the process of linking empirical data.

\section{What compose a basic physical quantity?}\label{addition}

Any teacher concerned with instilling the right attitude toward mathematics among physics students has at least the following thought to convey to them: In contrast to such numbers as, for example, the Mohs hardness of solids or the Mercalli intensity of earthquakes, those present in the mathematical expressions of laws of Nature have no option to be defined arbitrarily or by convention. While the former, termed ordinals in mathematics but often referred to as ranks or ranking numbers, are nothing but the labels for linearly ordered data selected by human beings from observatins/experiments, the latter, termed cardinals in mathematics but usually referred to as quantities, constitute the set that natural regularities enable researchers to equip with algebra operations.

It is well known that in physics, aside the dimensionless numbers, we are allowed to add together only quantities of the same kind. Some of the addition operations, such as those applied to the masses of joined bodies, the forces of combined actions, or the consecutive displacements of a body, represent the basic physical regularities and reveal so distinctive features of the quantities involved that these operations can be called \textit{addition rules} for these quantities.
The necessary part of each addition rule is the non-numerical description of the experimental/observational relation between the physical objects/processes whose numerical magnitudes make the summands and the sum in the addition rule's formula. It seems reasonable to refer to this description as the \textit{composition}.

For example, in geometry \(l(AC)=l(AB)+l(BC)\) is the formula of the addition rule for the length \(l(MN)\) of an oriented line segment \(MN\) while the composition specifies that the union of the adjacent segments \(AB\) and \(BC\) of the same straight line is the segment \(AC\). Since a translation applied to one of two non-adjacent segments of a straight line can always transform them into two adjacent segments, it extends the concept of the composition to any two segments of a straight line.

Aiming to facilitate the process of accumulating new data in physics and engineering the university mathematics courses usually present geometry with the aid of Cartesian coordinates, which provides the description of translations of a point object as shift transformations of its coordinates. Within the process of linking the empirical patterns of spatial arrangements of physical bodies into explicit and implicit postulates of synthetic geometry, the translational movement of a body is a basic concept which logically precedes and begets those of body's dimensions and coordinates. To illustrate this point, we need go further and convert the notions in the above geometrical example, relating to segments of one straight line, into the almost physical ones.

Studying the set of all segments of a given straight line experimentally on a sheet of paper, a reseacher/learner can exploit some tools -- such as a ruler and a compass -- in order "to move" (i.e. to copy and put) any segment in any direction/place. But in contrast to the common courses on synthetic geometry, physicists cannot state in advance that the motions applied to the construction tools and/or to the segments themselves as physical bodies are "rigid." And, naturally, in a fully consistent analysis, physicists cannot postulate the existence of any unlimitedly fine instruments. Instead, their only research technique is to formulate some hypothetical conditions and then to verify experimentally whether they are satisfied or not by the available methods for setting the real physical bodies in motion.

Though the terms and notations in the initial example, such as the points \(A, B, C\), no doubt sound familiar to the reader, it is not good practice in the theoretical analysis to invoke more notions than necessary. For this reason, below in this section not geometric points but line segments themselves are considered to be basic objects. Ultimately, the right choice of the basic/primitive notions in a given theoretical construction facilitates the expansion of this construction, which is usually made by setting new basics/primitives and turning the old ones into the derived notions. In physics, any successful theory is strongly (though not always explicitly stated as) coupled with the class of feasible experiments or observable phenomena it describes. In particular, for many practical and research purposes, we can associate Euclidean geometry with the variety of possible \textit{mutual arrangements} of sufficiently rigid macroscopic bodies. This also includes drawings on a sheet of paper, which serve so well for the purposes of teaching. By convention, the other types of experimental/observational data, such as those on \textit{relative motions} of the bodies as well as on the properties of their materials, let alone electromagnetic/optic phenomena, are just the subjects of physics itself.

Note that before any attempts of studying the real physical world a researcher or a learner faces only the raw data, i.e. the set of unclassified, unstructured, unrelated things and events. The reader can assume that the most coarse classification of such data simply corresponds to one or another historical division of science into its branches; only the ability to distinguish between geometric and non-geometric facts to some extent is important below. (This data classification is a starting point of the inference designed to provide further specification. Similar to an initial guess in the iterative computation method, it is allowed to be somewhat arbitrary.) The rest of the section illustrates how a physicist could (and should) make geometrical data linked, up to defining the length of a segment as a physical quantity.

To begin the process of building the desired mathematical structure, we should start to refer to two bodies as geometrically equal/coinciding if they are indistinguishable without addressing their non-geometric properties. The following definition extends the concept of the geometric equality to two segments at different places. Two segments \(a\) and \(b\) of the straight line can be called congruent, \(a\approxeq b\), if
\begin{enumerate}
 \item a longitudinal motion of either \(a\) or \(b\) along the straight line juxtaposes them so that they appear equal; 
 \item this relation is symmetric, i.e.
       \begin{equation}\label{segment symmetry condition}
              a\approxeq b \text{ implies } b\approxeq a
       \end{equation}
       for any two segments \(a\) and \(b\), and transitive, i.e.
       \begin{equation}\label{segment transitivity condition}
              a\approxeq b \text{ and } b\approxeq c \text{ imply } a\approxeq c
       \end{equation}
       for any three segments \(a\), \(b\), and \(c\).
\end{enumerate}

If one changes the original relation 'being congruent' into 'being either the same segment or congruent', one finds this modified version automatically symmetric and transitive and, additionally, reflexive, i.e.
\begin{equation}\label{segment reflexivity condition}
    a\approxeq a.
\end{equation}
In mathematics, such a binary relation defined between any two elements of a given set is called an equivalence relation (in this set).\cite{equivalence relation} In physics, it is reasonable to apply this term also to a relation defined between any two \textit{different} elements of that set and satisfying only \eqref{segment symmetry condition} and \eqref{segment transitivity condition} since the trivial modification, described above, automatically extends it up to that compliant with the standard formulation.

The entities being in a given equivalence relation are generally referred to as equivalent (with respect to that relation). The equivalence relation in a given set determines a partition of that set into its pairwise disjoint subsets containing only equivalent elements and usually referred to as equivalence classes of that set.\cite{partition theorem} More specific terms are possible for each specific equivalence relation. For example, we can say that the congruence relation begets the congruence classes.

The requirement that congruence is an equivalence relation imposes restrictions on the possible uncertainty associated with the notion of line segment but can not eliminate it completely. Most of the readers have no doubt been taught to describe the uncertainty in observable physical events or in measurable physical regularities only by means of an absolute error (along with its statistical estimation known as standard deviation) and a relative error, which both eventually refer to some values of the physical quantity relevant to the phenomenon under study. Historically, the concept of absolute error gave birth to the fundamental construct of real analysis, the converging sequence of real numbers, since such a list can be conceived as the abstraction of successive approximations of some quantity in attempt to make their absolute error less than a given value. However, when a researcher explores natural phenomena at the level that can not support the existence of physical quantities as real numbers, the above familiar tools for monitoring the accuracy of the experimental and/or theoretical investigation are not applicable. Since this essay deliberately focuses on that level, the following comment about the more general description of accuracy will be helpful.

To make things easier, one can limit the consideration to the set of all segments within a given finite segment of the straight line, below referred to as the base segment. Then the uncertainty in the congruence relation manifests itself in the \textit{finite} number \(r\) of the congruence classes in that set. Clearly, all sufficiently small segments have to make one class. Some larger segments make next class and so on. We can call \(r\) the congruence resolution (in the base segment).\label{resolution} An increase in this number is the sign of improving the accuracy of congruence: The change \(r\to r'=r+1\) means that a researcher succeeds in discerning more congruence classes. Note that only the difference of two resolutions is essential while the actual value of \(r\) itself need not even be observable, especially in case of a sufficiently large base segment. Also, it is important to understand that at this stage there is no way to compare the accuracy improvements \(r_1\to r_1'=r_1+1\) and \(r_2\to  r_2'=r_2+1\) for \(r_1\ne r_2\) since no characterization for the "size" of a congruence class has been introduced yet. This means that, though by definition \(r\) is a natural number, when describing the accuracy it represents only an ordinal number.

The concept of congruence makes movable segments of a straight line similar to rigid physical bodies. The concatenation \(a\cup b\) of two segments \(a\) and \(b\) as their composition seems compatible with this picture. Still, there remains to be regularities to establish/verify on the way to the concept of length. For example, one should make sure that for each segment \(a\) there is such a quantity \(l(a)\) that
\begin{equation}\label{segment commutation}
    l(a)+l(b)\equiv l(a\cup b)=l(b\cup a)\equiv l(b)+l(a).
\end{equation}
Someone with an engineering background may wonder: "When I close the jaws of the slide caliper around a mechanical part of interest and count, say, 7 contiguous spacings of 1 mm in the caliper's graduation from its origin mark to the pointer's position, doesn't it mean that there is a dimension of this part which is congruent to the composition of 7 standard line segments we simply agree to call 'millimeter'? If it does, then don't the equalities like 3 mm + 4 mm = 4 mm + 3 mm follow from nothing but the axioms (i.e. the definitional description) of natural numbers and simply mean that I can count any objects in any order and so there is no need to check Eq. \eqref{segment commutation}?"

In response to these questions, an experienced theorist could explain: "The dimension of a macroscopic object in a given direction does exist with accuracy sufficient for engineering applications, though not due to an act of measurement but due to a number of physical regularities that govern the behavior of macroscopic solid bodies and, for this reason, underlie various procedures of measuring length. Even if we are able to recognize what segments are congruent, Eq. \eqref{segment commutation} still needs to be checked experimentally since it hinges on another notion, a composition of two segments. Physically, this notion denotes the result of fastening one body to another so that the joined bodies can be set in motion without breaking their bond. The idea that the composition has its origin in the real physical interaction entails some reservations about this notion and, therefore, limits its application area. For example, when one attaches a body \(a\) to a body \(b\), one expects that such an act makes no effect on the spatial extent of both \(a\) and \(b\). However, this presumption may practically turn out to be questionable for either measuring unusually small microscopic distances with usual accuracy or measuring usual macroscopic dimensions with unusually high accuracy. In other words, a researcher should be prepared to find no sufficiently rigid body that he/she could call 'nanometer'.
Then the ability of physics to maintain the concept of length, i.e. the statements like Eq. \eqref{segment commutation}, doesn't look so trivial."

Thus, the necessary logical step toward the definition of length is to state explicitly that any \(a\approxeq a_1\) and \(b\approxeq b_1\) entail \(a\cup b\approxeq a_1\cup b_1\) where the symbol \(\cup\) denotes contact interaction between the end of one segment and the end of another segment. As soon as this statement is experimentally verified one can use \(a\cup b\) and \(a_1\cup b_1\) interchangeably to simplify the notation of more complex congruence relations. 

Then one should make sure that the relations
\begin{equation}\label{composition commutativity}
     a\cup b\approxeq b\cup a
\end{equation}
and
\begin{equation}\label{composition associativity}
      (a\cup b) \cup c\approxeq a\cup (b\cup c)
\end{equation}
are realizable physically within a given resolution for any three segments \(a\), \(b\) and \(c\). Here a pair of brackets is used to tag a composition treated as one segment in the sense of all previous statements. Note that fact that the relation \eqref{composition commutativity} itself entails the congruence of some compositions of more than two segments, e.g. \(  (a\cup b)\cup c\approxeq c\cup (b\cup a) \) while the relation \eqref{composition associativity} is an independent statement.

As far as the above conditions are met, one can think of the line segments as rigid rods. In particular, the relations \eqref{composition commutativity} and \eqref{composition associativity} enable us to use the abbreviated (and more evocative) notation \({}^na\) for the composition of \(n\) segments each congruent to a given segment \(a\) because the order of the segments does not matter now. If \({}^na\approxeq {}^mb\) with a pair of natural numbers \(n\) and \(m\) then it defines the ratio
\begin{equation}\label{ratio}
      R(a|b)=m/n.
\end{equation}
Evidently, the rational number \(l(a)\equiv R(a|s)\) is simply what we agree to call the length of the segment \(a\) when accepting \(s\) as a standard segment.

This section exploits the line segments to show how the concepts of equivalence relation and composition can lead to the existence of a physical quantity, the length of a segment. However, since any hint of the physical definition of a straight line is beyond the scope of this essay, a line segment here is only a toy example of a physical object. The next section outlines the pre-numeric introduction of a well known quantity applicable to the real physical bodies.

\section{Which came first, the weight or the balance?}\label{weight}

Section \ref{perplexing science} makes it clear that the presentation in common textbooks as well as, inevitably, in their respective courses hinges on a huge logical leap from the everyday notions to the vaguely defined abstract concepts, rendered meaningful only by subsequent examples. In view of this practice, it would be no surprise to hear from a university graduate in physics/engineering that to design, to assemble and then to adjust a beam balance requires the prerequisite theoretical knowledge such as geometry, the concepts of the mass and the weight, the lever principle and perhaps even the elasticity/plasticity theory to bring the measurement errors under control as completely as possible.

However, there are hardly any doubts that there were no theoretical concepts involved in the invention of such an ancient device. It is the direct juxtaposition of raw experimental data that any experimenter has recourse to when the theory relevant to his research is yet to be developed. Along the way to the discovery of the mass as a quantity, these preliminary observations may have been carried out over a bar that has one hinge as a pivot and two hinges to hold weights. The simple comparison of data obtained with different bars must have led observant people to the conclusion that a certain group of such bars can exhibit a \textit{consistent} behavior: for any pair of weights (within the scope of the bars' sensitivity; see below), balancing each other on one bar implies doing so on any other bar. Then each bar of that group became acceptable as a device named a balance. Naturally, the bar-like shape is not essential for a device to be a balance. Even sophisticated intelligent system will do as long as its use results in the same relation between weights as that of other balances. But there must be at least several reference balances as simple as identifiable chunks of solid matter. Pivot's motion and orientation may have an effect on the balance, yet adhering to the consistent observations is enough for a researcher/learner to recognize and avoid incorrect use of the balance.\cite{equivalence principle} This consistency makes the relations between weights, further described later on in this section, independent of the choice of a balance within some group of such devices.
 
At this early stage of the theoretical development there would not yet have been any numbers to characterize each equilibrium state of a balance loaded with weights. So a researcher would have no choice but to mark each of the bodies under consideration and then simply classify (the relations between) them. This would apply to all aspects of such a study, including the description of accuracy. In particular, for each equilibrium state of a balance there could be some groups of sufficiently light bodies the addition of which to one of the balance's scalepan don't affect the equilibrium due to friction at the fulcrum. Then, for want of better way to characterize the sensitivity of a given balance, the researcher would have to enlist bodies that make up each of those groups. 

Section \ref{addition} shows how comparison and composition properties give rise to the existence of a physical quantity. It is not difficult to see that placing two bodies in the opposite scales of a balance and then observing its final state is just a comparison procedure analogous to juxtaposing two line segments. So this procedure applied to each pair of bodies from a set of physical bodies results in the ordering of this set. In an analogy to the concept of congruence, the researcher/learner can use the balance to verify that the binary relation of 'being equiponderant', hereinafter denoted as "\(\approxeq\)," is an equivalence relation in a set of physical bodies and therefore partitions this set into the classes of equiponderant bodies. It is worthwhile to keep in mind that in this context finding a real body equiponderant to a given body relies on trial and error, which is the most basic technique available to researchers.

The equiponderance relation makes it possible to introduce the equiponderance resolution. The definition for this simplest structured characterization of the accuracy of equilibrium within a given set of bodies is similar to the congruence resolution in the geometry of line segments (see p.\pageref{resolution} in Section \ref{addition}). In the practice of weighing macroscopic bodies the uncertainty in equilibrium is almost entirely due to the limited sensitivity of a balance. (Since, by definition, a balance belongs to a group of mutually consistent devices, one can also say that all the devices in the group share the same sensitivity.) Because of this, the concept of the equiponderance resolution can be directly demonstrated by a teacher in the classroom or experienced by students in their lab activity. To do so, they should choose a certain amount of sand (or other easily divisible substance) and then make all those pairs of equiponderant portions of the sand that are distinguishable from each other with an aid of a given balance. Specifically, they can start with equipoise of the two halves of the initial amount of the sand on the two pans of the balance, below referred to as the pan 1 and the pan 2. The experimenters should make sure that removing sand from the pan 1 does not disturb the balance until the amount of the removed sand does not exceed a certain threshold. Then the various amounts of the remaining sand in the pan 1 that can keep equilibrium with the untouched sand in the pan 2 make the first equiponderance class. As soon as the gradual removal of sand in the pan 1 starts breaking equilibrium, the experimenters should swap the roles of the pan 1 and the pan 2. Then the varied amount of sand in the pan 2 that can keep equilibrium with the fixed amount of sand in the pan 1 represents the second equiponderance class. If one repeats these procedures until the exhaustion of sand in the pans, one finds the total number of the equiponderance classes, i.e. the equiponderance resolution of the balance with respect to the initial amount of sand.

Besides heaps of sand, there are other bodies whose heavinesses one can compare with a given balance. Naturally, the complete equiponderance classes of the balance are limited by its strength and dimensions. If an equiponderance class does include heaps of sand, the spread of the sand amounts in the heaps is a distinguishing characteristic for the whole class. So the "sand" equiponderance resolution, described above, is a simple gauge of uncertainty of the equiponderance, which logically precedes both the notion of composition of several bodies and the ensuing concept of mass/weight.

As the reader can see in Section \ref{addition}, the necessary step toward the definition of a basic physical quantity is to verify that the equivalence of two compositions follows the equivalence relations between their respective components. But whereas a line segment, the model presented in Section \ref{addition}, has only two ends where other segments of the same line can join to form another segment\cite{arrangement order}, the surface of a real solid has numerous sites where one can attach other solids to make a compound stable within a given class of its motions. So the reseacher/learner interested in the fully consistent introduction of the mass/weight should make sure that any \(a\approxeq a_1\) and any \(b\approxeq b_1\) together entail \(a\cup_i b\approxeq a_1\cup_j b_1\) for any two bonds \(\cup_i\) and \(\cup_j\) between two bodies \(a\) and \(b\) and between their counterparts \(a_1\) and \(b_1\). Such a statement implies that \(a\cup_i b\approxeq a\cup_j b\). It makes all bonds between two given bodies equivalent, so one can use the symbol \(\cup\) to denote the presence of a bond without specifying its type. Under the current meaning of all the symbols as they are used in this section, the relation \eqref{composition commutativity} holds automatically while the relation \eqref{composition associativity} does need verifying. It is after completing this verification (for a given set of bodies and for a given equiponderance resolution), one can assert the existence of the rational numbers \eqref{ratio} and then define the mass \(m[a]\equiv R(a|s)\) of a body \(a\) with the aid of a standard body \(s\). Here a 'standard body' means a reference body that represents only itself. Meanwhile, nothing prevents us from adapting a reference body to represent a standard action, for instance, a certain extension/contraction of a certain spring loaded by this reference body. If this is done under the same conditions in which one uses the balance to weigh a given body, then weighing will actually yield the quantity conventionally called the (apparent) weight of this body.\cite{ambiguity}

Those who apprehend the logic of presentation in this and previous section may guess how one could further proceed with introducing basic physical quantities: For instance, the consistency of balances, outlined in the beginning of this section, can be described accurately in terms of an equivalence relation between two balances. By combining it with some rules for the composition of balances one can eventually come to another well known quantity. The same applies to the bonds between parts of solid and further concepts that a researcher/learner would inevitably refine while getting increasingly detailed description of interactions between macroscopic bodies. However, the space limitations for a journal publication does not allow this essay to extend that far. Instead, in the next and final section the reader can find some suggestions of how teachers can take advantage of the logically consistent presentation of basic physical quantities.

\section{What is the benefit?}

The main concern of this essay is to show how and why the pre-numeric concepts such as an equivalence relation between and a composition of physical entities give rise to a numeric characterization of these entities. This is exemplified by the logical chain of introducing the mass/weight of a physical body. Quite naturally, to deliver this presentation verbatim in the classroom can hardly help a teacher to fulfil the traditional objectives of physics classes, especially those for non-physics majors. Nevertheless, the mere existence of the properly ordered hierarchy of the basic physical concepts can immediately benefit at least the following three groups of teachers: 

The teachers from the first group are young instructors that claim no significant expertise in curriculum development but feel obliged to be representatives of science in the classroom. They would like to help their students in getting knowledge from standard textbooks, yet it irks them to take responsibility for the textbooks' logic. Evidently, it is unlikely that the students would be able to question this logic by themselves. However, the slipshod treatment of the fundamentals, such as outlined in sections \ref{foundationless}-\ref{logic of science}, is the very deficiency in the standard texts that inclines the increasing number of unqualified people to perceive physical theories on par with pseudo-scientific arbitrary constructs.  As a result, there are many web forums, blogs and other sites where they comment on various aspect of science presented in education and in public media. This makes it possible for an unexperienced teacher to face the embarrassing questions, which he/she cannot ignore without losing his/her credibility in the classroom. Now, in response to questioning the mandatory texts, he/she can soundly explain that there is a technique to introduce the basic concepts seamlessly from experiments if need be but it goes beyond the current curriculum.

The second group consists of the teachers that have enough experience to understand that real physics is an experiment-based science, which can adopt theory only to the extent justified by experimentally driven problems. So they never see solid reasons to worry about such thing as the poor presentation of theoretical notions. Still, they probably favor those minor changes in physics courses that would make experimentation more proactive in expounding fundamentals rather than focused on the passive choice between several externally devised hypotheses. The technique outlined in this essay just shows how to organize the initial trial and error process of discovering physical regularities within the most rational, purely phenomenological approach.

The teachers/lecturers from the third group are supposed to be both theoretically-minded and bold/authoritative enough to update the physics curriculum. They can take advantage of the pre-numerical level of theoretical description, such as one used in Section \ref{weight} or more elaborated, to rebuild the beginning of their physics course in a theoretically impeccable manner. Obviously, the curricula for physics majors need to be changed first. But even as early as in the middle school science lessons, it would also seem natural to address some of the pre-numerical notions as a first acquaintance with the theory underlying the measurement procedures, instead of premature introduction to the algebra-based concepts of Newtonian mechanics.

Successful integration of the pre-numerical formalism into physics curricula for the introduction of even one physical quantity will turn it into a model for a theoretically consistent and experimentally well-grounded presentation, which will then stimulate improving other topics in the curricula and developing higher standards for the explication of physical phenomena in both educational and research activities.

\appendix

\section{Do you know how to subtract?}\label{subtraction examples}

To get the point of the heading across to a person with the middle school level of education I recommend him to assess the validity of using the minus (subtraction) sign "\(-\)" in the following entries:
\begin{equation}\label{seemingly wrong}
  15 \text{ pigs } - 10 \text{ ducks},
\end{equation}
\begin{equation}\label{seemingly dubious}
   10 \text{ inches in length } - 2 \text{ inches in width},
\end{equation}
\begin{multline}\label{seemingly right}
	5 \text{ meters ahead at } 3\text{rd second} \\ 
	- 2 \text{ meters left at } 1\text{st second}. 
\end{multline}
I then ask the student to interpret/impove these expressions to make them meaningful.

"What a nonsense! If someone has nothing but fifteen pigs at his disposal how could he find among them ten ducks to remove? Anyway, I was taught you cannot add quantities of different kinds together, nor subtract one from another." This would be a likely response to the entry \eqref{seemingly wrong}.  And I would make the following comment to it: "You were taught about using arithmetics quit correctly. But perhaps that wouldn't be the first thing that came to your mind if I drew your attention to the similar entry '\( 10 \text{ pigs } + 3 \text{ ducks }\).' Apparently, this is because there is no problem to put three ducks among ten pigs for the time being. After some thought, you will be able to come up with a more or less plausible interpretation of the entry \eqref{seemingly wrong}, too.  Say, it might be a note of a farm owner where the shorthand '\(-\)' indicates a debt to his fellow farmer. One could also imagine that those animals are actually waxworks, so that the entry \eqref{seemingly wrong} might involve melting some of the figures of pigs to cast the figures of ducks from the resulting wax. Note that the subtraction rule in this last example necessarily includes the knowledge about the nature of the objects to which it applies. The same holds for the basic physical quantities in a standard introductory physics course, because all relevant addition and subtraction operations are possible only in accordance with the formulas that express the physical laws relating such quantities."

This preliminary discussion encourages the student to get more careful and to suspect that the entry \eqref{seemingly dubious} and maybe even the entry \eqref{seemingly right} need improvement since, at least formally, both look as if the subtraction is applied to the quantities of different types. Now, interpreting the entry \eqref{seemingly dubious}, the student will be able to come to the idea that somebody cuts a square from one end of a paper strip 10 inches long and 2 inches wide. Then the student's corrected version of that entry will probably look like
\begin{multline*}\label{seemingly dubious improved}
 	10 \text{ inches in length } - 2 \text{ inches in length obtained}\\
	\text{by rotation from } 2 \text{ inches in width}
\end{multline*}
Thinking about the entry \eqref{seemingly right} in its turn, the student will guess that the moments of time, which seem additionally labeling everything, are a characterization of an external construct named a reference frame whose construction/motion may affect the result of the subtraction in  the entry \eqref{seemingly right}.

\section{Do we know our limits?}\label{limitation}

In accordance with the factual empirical basis of Newton's laws, one should provisionally define classical mechanics as a theoretical description of interactions and relative motions of (the parts of) macroscopic bodies. Aside from such regularities as Hooke's law or Newton's law of universal gravitation, the theory should necessarily include Euclidean geometry as a description of the properties of the potentially possible arrangements of rigid solids.

Due to the inevitably limited accuracy of their empirical sources, the various components of the theory are not compatible entirely. This essay is not the place to analyze the logical structure of classical mechanics in detail and to evaluate the degree of its inconsistencies in the usual form of the relative error dependent on parameters of a particular problem. Still, there are situations where an initial assumption of the theory fails apparently, so it is not difficult to estimate the marginal value of a parameter for which the relative error of the theory is of the order of unity. Below the reader can see how to find out the lower bound \(a\) of the spatial scales (or, equivalently, the size of a part of a macroscopical body) over which classical theory is applicable.

The important thing that classical mechanics implies about a macroscopic body is that its material is able to convey the external force from its one part to another. Newton employed this empirical fact in his \textit{Principia} to prove his third Law of motion for the long-range interaction (see p.~\pageref{III Law} in Section \ref{logic of science} of this essay.). Meanwhile, it is common engineering knowledge that any material can withstand a tensile or compressive stress that does not exceed a certain value called the strength of the material. Although fewer imperfections in the crystal structure of solid or less impurities in liquid entail its higher strength, each such quantity ultimately appears to be finite and thereby sets a limit for Newtonian mechanics.

For teaching/analyzing this limitation of classical mechanics the most simple relevant experimental data available are on the cavitation in liquids stretched by a negative pressure \(-p\). In other words, a teacher/theorist who gives due attention to the process of \textit{linking} empirical data (see p.~\pageref{science activity} of this essay) has little choice but to turn to what makes such liquid unstable: a spherical cavity that can grow. If the surface tension of the liquid is \(\sigma\), then the minimal radius of such cavity,
\begin{equation}\label{equilibrium cavity radius}
    a\approx\frac{2\sigma}{p},
\end{equation}
follows from the hydrostatic equilibrium condition (see, e.g., Eq. (1.15) at p.~12 in Ref.~\onlinecite{Skripov1974} where the vapor pressure \(p''\) can be neglected). In the following, it is assumed that \(p\) has reached its bound, i.e. equals the tensile strength of the liquid. Then Eq. \eqref{equilibrium cavity radius} also tells us about the maximal radius of a cavity arising \textit{spontaneously} in the stretched liquid. And for the sake of consistency, in pure liquid, we should ascribe the origin of such an initial cavity to deviations from the classical laws. Thus \(a\) turns out to be an estimation for the desired marginal size.

Someone may object that the above reasoning relies on an extrapolation of classical mechanics (specifically, hydrostatics) laws to the spatial scale where they are not supposed to be applicable. In fact, this extrapolation is a part of the definition of the marginal spatial scale \(a\), at which classical and non-classical effects are expected to be of the same order of magnitude. To make the reasoning formally compatible with non-classical contribution/modification one should simply use the symbol "\(\simeq\)" in Eq. \eqref{equilibrium cavity radius} in place of the symbol "\(\approx\)." Note that this enables us to refer to the estimation based on Eq. \eqref{equilibrium cavity radius} as empirical, i.e. as true as the input empirical data.

For water \(p\simeq 30\) MPa and \(\sigma\approx 70\) mN/m at the temperature about \(20 {}^\circ\)C (see p.~4 in Ref.~\onlinecite{Caupin2012} or p.~041603-17 in Ref.~\onlinecite{Herbert2006}) so one can find \(a\simeq 5\) nm. The dynamical timescale \(\tau\simeq 2a/c_s\approx 7\) ps of the corresponding disturbances (where \(c_s\approx 1.5\) km/s is the speed of sound in water) is a more informative characteristic of the cohesive interaction between parts of water. In particular, the interactions governing the spontaneous density fluctuations and the dissipation of the externally induced waves are the same, and so are their characteristic timescales, because something we treat classically as growing and collapsing bubbles underlies both the phenomena.
Note that the relaxation time \(\tau\approx 8 ps\) actually manifests itself in water (see  p.~197802-2 in Ref.~\onlinecite{Fukasawa2005}).

Data relating to liquid helium allow us to estimate \(a\) as the temperature approaches 0 K: The \({}^3\)He tensile strength 3 bar (see p.~S79 in Ref.~\onlinecite{Balibar2003} and  p.~64064507-1 in Ref.~\onlinecite{Caupin2001}) along with its surface tension 160 mdyn/cm (see p.~291 in Ref.~\onlinecite{Iino1985}) and the \({}^4\)He tensile strength 9 bar (see p.~S76 in Ref.~\onlinecite{Balibar2003} and  p.~64064507-1 in Ref.~\onlinecite{Caupin2001}) along with its surface tension \(375 \mu J/m^2 \) (see p.~565 in Ref.~\onlinecite{Roche1997}) yield \(a\simeq\) 1 nm for both \({}^3\)He and \({}^4\)He. Note that all the values of \(a\) far exceed a characteristic atomic spacing, the Bohr radius 0.053 nm.


\begin{thebibliography}{5}
\bibitem{casual definitions}The casual/provisional definitions, such as discussed in Ref.~\onlinecite{Galili2006}, may be useful as starting points of learning/research but can not be any part of established knowledge, the organization of which, rather than the pedagogy, is basically the subject of this paper.
\bibitem{amalgamation} Someone may also interpret it as identifying additional links between the previously established laws. Feynman called this process "amalgamation" (see p.~2-2 in Ref.~\onlinecite{Feynman1}).
\bibitem{equivalence principle} In a further presentation we could define such balances as non-rotating bodies and later, after the introduction of the concept of acceleration, identify them as (differently) accelerated ones. This order of presenation makes it redundant to distinguish between so called 'inertial' and '(passive) gravitational' mass since we address the equivalence principle at pre-numerical level, while selecting the group of the consistent devices.
\bibitem{arrangement order} The presence of two types of bonds between two oriented line segments in Section \ref{addition} is implicitly encoded in distinguishing the order of the two possible arragements of the segments in their composition. This is an unavoidable part of the formal definition of a segment that does not address the concept of the geometric point.
\end{thebibliography}
\end{document}